\documentclass[%
 reprint,
 amsmath,amssymb,
 aps,
]{revtex4-2}

\usepackage{steinmetz}
\usepackage{graphicx}
\usepackage{dcolumn}
\usepackage{bm}

\usepackage{changes}
\usepackage{amsmath}
\usepackage{comment}
\usepackage{color,soul}
\usepackage{hyperref}       
\hypersetup{
    colorlinks=true,
    linkcolor=blue,
    filecolor=magenta,      
    urlcolor=cyan,
}
\begin{document}

\preprint{APS/123-QED}
\title{Quiet point engineering for low-noise microwave generation with soliton microcombs}


\author{Andrea C. Triscari$^{1,2}$}
\author{Aleksandr Tusnin$^{1,2}$}
\author{Alexey Tikan$^{1,2}$}%
 \email{alexey.tikan@epfl.ch}
\author{Tobias J. Kippenberg$^{1,2}$}%
 \email{tobias.kippenberg@epfl.ch}
\affiliation{%
 $^1$Institute of Physics, Swiss Federal Institute of Technology Lausanne (EPFL), Lausanne, Switzerland\\
 $^2$Center for Quantum Science and Engineering, EPFL, Lausanne, Switzerland \\
}%

\date{\today}

\maketitle



\textbf{Low-noise microwave signals can be efficiently generated with microresonator-based dissipative Kerr solitons ('microcombs')\cite{kippenberg2018dissipative}.  However, the achieved phase noise in integrated microcombs is presently several orders of magnitude above the limit imposed by fundamental thermorefractive noise~\cite{liu2020PhotonicMicrowaveGeneration}. 
One of the major contributors to this additional noise is the pump laser frequency noise transduction to the soliton pulse repetition rate via the Raman self-frequency shift~\cite{yi2016TheoryMeasurementSoliton,karpov2016RamanSelfFrequencyShift}. Quiet points (QPs) allow minimizing the transduction of laser frequency noise to soliton group velocity~\cite{yi2017SinglemodeDispersiveWavesa}. While this method has allowed partial reduction of phase noise towards the fundamental thermodynamical limit, it relies on accidental mode crossings and only leads to very narrow regions of laser detuning where cancellation occurs~\cite{lucas2017DetuningdependentPropertiesDispersioninduced}, significantly narrower than the cavity linewidth. Here we present a method to deterministically engineer the QP, both in terms of its spectral width, and position, showing an increased phase noise suppression. This is achieved using coupled high-Q resonators~\cite{tikan2021emergent,tikan2022protected} arranged in the Vernier configuration~\cite{helgason2021dissipative}. Investigating a generalized Lugiato-Lefever equation~\cite{lugiato2018lugiato} that accounts for the hybridized mode spectral displacement, we discover a continuum of possible QPs within the soliton existence region, characterized by ultra-low noise performance.  Furthermore, we discover that by using two controlled optical mode crossings, it is possible to achieve regions where the QPs interact with each other enabling a substantial increase of the noise suppression range. Our work demonstrates a promising way to reach the fundamental limit of low-noise microwave generation in integrated microcombs.}

\section{Introduction}
The discovery of dissipative Kerr solitons (DKS)~\cite{herr2014temporal} in driven dissipative Kerr nonlinear resonators,  has heralded a new method to synthesize coherent and broadband optical frequency combs, with compact form factor, wafer scale manufacturing compatible techniques, and mode spacings that can access the microwave to THz. The dynamics of such DKS in microresonators is described by the Lugiato-Lefever equation (LLE)~\cite{lugiato2018lugiato}. It is by now well understood that microcombs give rise to a plethora of coherent nonlinear dynamical states, i.e. 'dissipative structures', including platicons, zero dispersion solitons, and solitons in unusual dispersion landscapes 
\cite{yu2021spontaneous,helgasonDissipativeSolitonsPhotonic2021}, that challenge the understanding of 'classical' bright solitons. DKSs can crucially be generated in photonic integrated microresonators based on silicon nitride \cite{Brasch2016-kp}, a foundry level, mature photonic integrated circuit technology, that has been the basis of numerous system level demonstrations,  including massively parallel~\cite{riemensberger2020massively}, dual-comb ~\cite{Lukashchuk2021Dual} and chaotic LiDARs~\cite{Lukashchuk2021Chaotic}, neuromorphic computing~\cite{Feldmann2021Parallel, xu202111}, as well as optical frequency synthesis~\cite{Spencer2018-oc} and optical clocks~\cite{Papp:14}
Microcombs also enable low-noise microwave generation by detecting the repetition rate of the soliton pulse stream. Such optically generated microwaves are attractive due to the low power and potentially low phase noise that can be generated and can be employed in a variety of applications, such as microwave photonics, Radar~\cite{Khan2010-uu}, 5G/6G~\cite{Lima2022-nk} or  wireless communications~\cite{rappaport2011state}. In contrast to optical frequency division, which employs phase stabilized femtosecond laser frequency combs, the phase noise of the generated microwaves in the case of microcombs is determined primarily by transduction of laser phase noise to the soliton group velocity. 
There have been numerous demonstrations of the soliton stream-based low-noise microwave generation
~\cite{savchenkov2008PhaseNoiseWhisperinga,papp2011SpectralTemporalCharacterization,li2012LowPumpPowerLowPhaseNoiseMicrowave,liang2015HighSpectralPurity,lucas2017DetuningdependentPropertiesDispersioninduced, yi2017SinglemodeDispersiveWavesa,Weng2019,stone2020HarnessingDispersionSoliton}, ranging from  sources in the K and X microwave band using integrated Si$_3$N$_4$ microresonators~\cite{liuPhotonicMicrowaveGeneration2020a} to the THz domain~\cite{Kuse2022-be}. Despite achieving phase noise on par with mid-range commercial microwave generators based on quartz oscillators, the fundamental limit of the repetition rate noise, as given by thermo-refractive noise (TRN) \cite{huang2019ThermorefractiveNoiseSiliconnitride,stone2020HarnessingDispersionSoliton,Yang2021}, is still several tens of decibels below the best experimentally demonstrated noise performance photonic chip-scale microcombs. While soliton microcombs have achieved lower noise, this has so far only been possible in bulk polished crystalline resonators, that have significantly lower TNR levels, and do not show Raman self-frequency shifts \cite{Lucas2020}. In contrast, for chip-integrated microcomb platforms, such as based e.g. on silica, silicon nitride, etc. the presence of the Raman self-frequency shift-related transduction noise limits  
achieving the fundamental thermodynamical limit~\cite{huang2019ThermorefractiveNoiseSiliconnitride,stone2020HarnessingDispersionSoliton,Yang2021}.
One way to reduce this noise has been the observation of 'quiet points~\cite{ yiSinglemodeDispersiveWaves2017a,lucas2017DetuningdependentPropertiesDispersioninduced}, which exploit a cancellation due to the interplay of the Raman self-frequency shift and the recoil from dispersive waves (DW) due to mode crossings (AMX).
\begin{figure*}
    \centering
    \includegraphics[width=\linewidth]{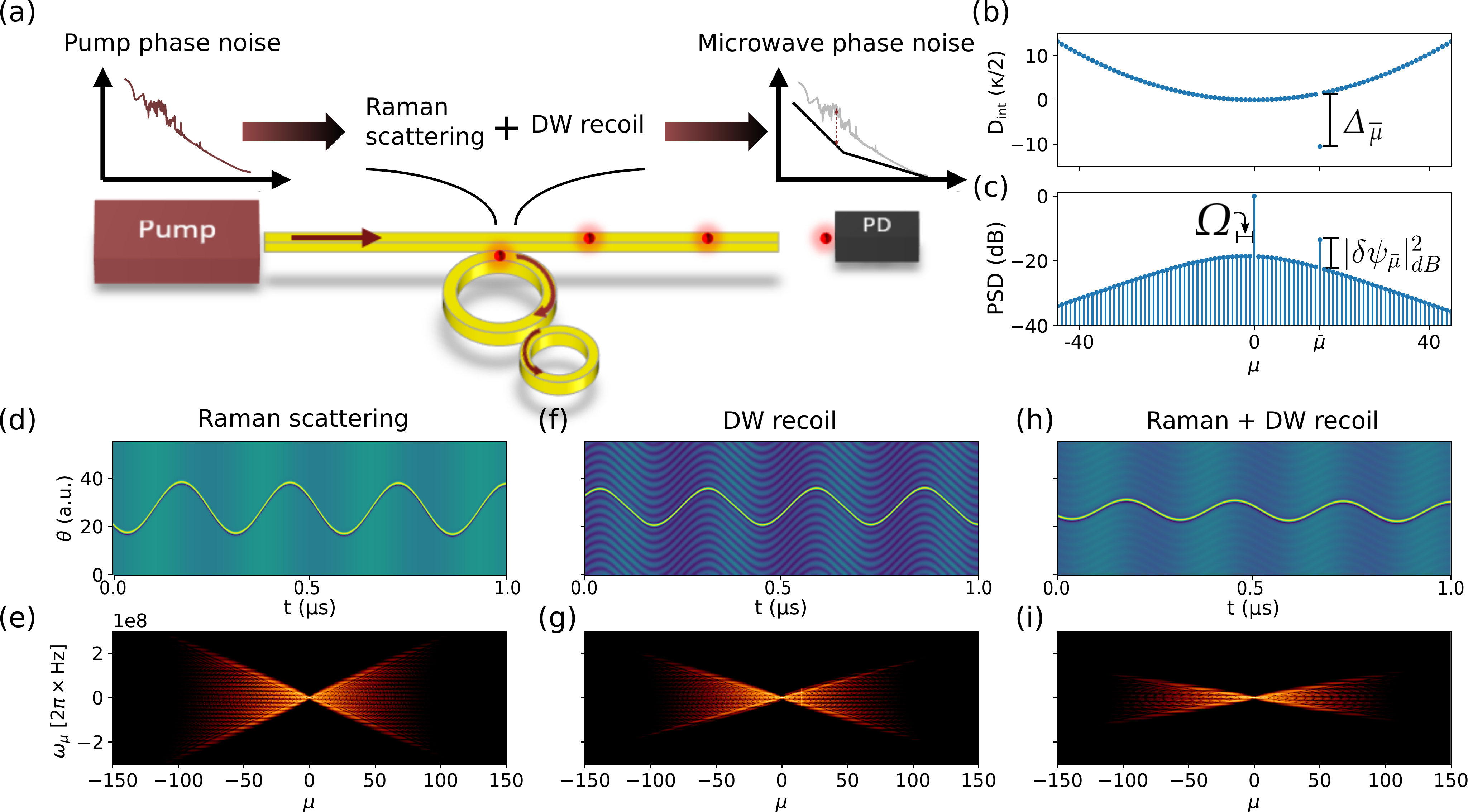}
   \caption{\textbf{The concept of the quiet point engineering using coupled resonators}. (a) Schematic of the phase-noise reduction in a microresonator. The auxiliary cavity is required to control the dispersive wave (DW) recoil in order to balance the effect of Raman scattering. (b) Example of the considered integrated dispersion profile with shifted resonance for a single spatial mode (i.e. localized AMX); the shifted mode number is $\bar{\mu} = 15$ and the resonance is shifted from the perfect parabolic profile by  $\Delta_{\bar{\mu}} = -4$. (c) Power spectral density of stable soliton solution both in the presence of Raman scattering and AMX for the integrated dispersion given in (b); the presence of Raman scattering and AMX results in a detuning dependent frequency shift $\Omega$ of the soliton spectrum responsible for the noise transduction mechanism. The shift induced by AMX on the dispersion profile leads to a perturbation $\delta \psi_{\bar{\mu}}$ of the occupancy of the displaced mode $\bar{\mu}$ with respect to the perfect hyperbolic secant profile resulting in the generation of dispersive waves. (d-e) Real-space (d) and Fourier-space (e) soliton dynamics driven with a  sinusoidal detuning ($\zeta_0(t) = 4 + 0.4 \operatorname{cos}(\alpha t + \varphi), \alpha = 0.0033, \varphi = -200 $) in presence of Raman scattering only (left), AMX (center) and both (right) for the dispersion profile in (b).}
    \label{fig:pics1}
\end{figure*}
However, this technique relies on accidental mode crossings and therefore is fixed to a certain mode number $\mu$ and shift $\Delta_\mu$ by design. Moreover, the strength of the mode crossing is a fixed parameter that is not subject to control~\cite{lucas2017DetuningdependentPropertiesDispersioninduced,yi2017SinglemodeDispersiveWavesa}. Moreover, the width of the QP has been reported to be substantially narrower than the linewidth of the microresonator cavity, implying possible second-order transduction effects.

In this work, we report on a deterministic approach to QP engineering and demonstrate a substantial increase in noise suppression bandwidth. We predict that with such improvements it is possible to reach the fundamental thermodynamical limit of phase noise, and thereby substantially improve the ability to synthesize low-noise microwaves directly from optical integrated microresonators. Our results are directly realizable within currently demonstrated silicon nitride integrated microresonator technology. 

\begin{figure*}
    \centering
    \includegraphics[width=\linewidth]{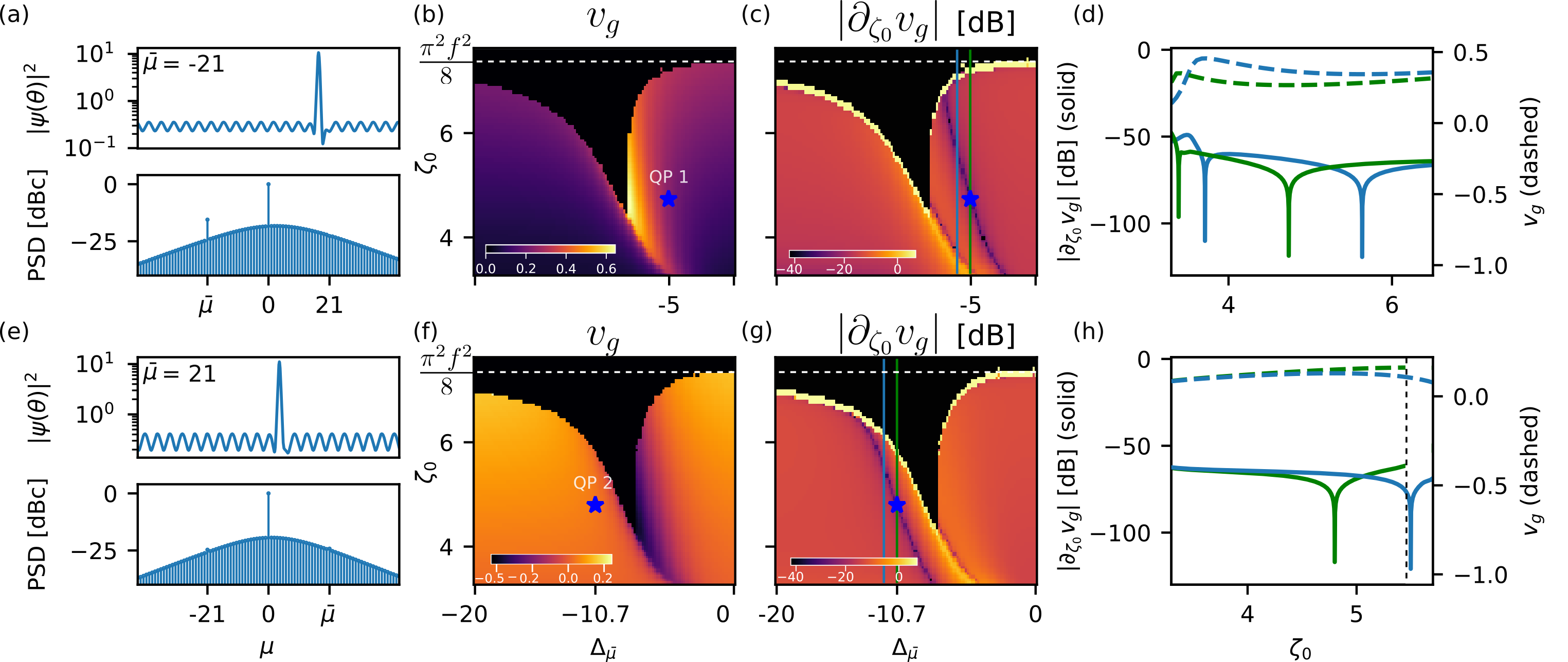}
    \caption{ \textbf{Quiet point identification via group velocity detection and dynamics in the 2D space of detuning-mode crossing strength}
    (a-d) Simulation results obtained with the Newton-Raphson algorithm for $(\Delta_{\bar{\mu}},\zeta_0;\bar{\mu} = -21, f^2 = 6, \tau = 5\cdot 10^{-3})$, i.e. in presence of Raman scattering and single mode resonance shift. a) Single soliton solution for $(\Delta_{\bar{\mu}} = -5,\zeta_0 = 4.73 )$ (blue star). The single mode shift leads to a periodic modulation of the constant background with period $\frac{2\pi}{|\bar{\mu}|}$. (b) single soliton group velocity $v_g$ in $(\Delta_{\bar{\mu}},\zeta_0)$-plane. The black area indicates the region of parameters where the method does not converge. The white dashed line highlights the existence range of the single DKS solution. The colormap shows how the soliton existence range is reduced in the presence of a higher value of the group velocity related to the increasing intensity of the DW that perturbs the DKS state. (c) $10\operatorname{log}_{10}|\frac{\partial v_g}{\partial \zeta_0}|$ as a measure of the DKS repetition rate susceptibility to the variation of  laser detuning. The smallest value of the susceptibility is obtained for a continuous line of operating point in the $(\Delta_{\bar{\mu}},\zeta_0)$-plane, here called the QP line (dark blue). (d) Comparison of the susceptibility (solid) and group velocity (dashed) profiles for two different sections of the QP line [green and light blue lines in subplot (c)].
    (e-h) Similar results for opposite mode displacement, i.e. $(\Delta_{\bar{\mu}},\zeta_0;\bar{\mu} = 21, f^2 = 6, \tau = 5\cdot 10^{-3})$.
    }
    \label{fig:pics2}
\end{figure*}

\section{Results}

To understand the transduction of phase noise to the soliton (i.e. DKS repetition rate $\omega_\mathrm{rep}$) we
consider Raman scattering and the DW recoil as the main noise transfer mechanisms (cf Fig.~\ref{fig:pics1}a. ), and aim to reduce the repetition rate susceptibility to the laser detuning fluctuations $\delta\omega$, i.e., minimize $|\frac{\partial \omega_\mathrm{rep}}{\partial \delta\omega}| \propto | \frac{\partial}{\partial \delta\omega}(\Omega_\mathrm{Raman} + \Omega_\mathrm{DW})|$~\cite{yi2017SinglemodeDispersiveWavesa}. To introduce the presence of a DW, we consider a simplified model of AMX, characterized by a single mode displacement at position $\bar{\mu}$ and strength $\Delta_{\bar{\mu}}$, in the integrated dispersion profile, as shown in Fig.~\ref{fig:pics1}b. As expected, the corresponding spectrum of the generated DKS has a typical $\operatorname{sech}^2$ shape,  frequency shifted by $\Omega$ and with spectral enhancement at $\bar{\mu}$, due to the DW  (Fig.~\ref{fig:pics1}c).     
To qualitatively explain the noise reduction mechanism, we start by separately analyzing  the Raman and DW contributions to the DKS repetition rate response. We consider the simplest case of sinusoidal frequency modulation of the pump detuning around a constant value $\delta\omega$ as presented in Fig.~\ref{fig:pics1}d-i. 
 In the presence of the Raman effect only, the DKS's group velocity is in-phase with the detuning change (c.f. Fig.~\ref{fig:pics1}d). In the nonlinear dispersion relation (NDR) representation~\cite{tikan2022nonlinear}, the DKS dynamics has a butterfly-shaped profile, revealing the transfer of the laser detuning modulation to the soliton group velocity $v_g$ (i.e., the tilt of the soliton line in Fig.~\ref{fig:pics1}e) that directly reflects the repetition rate change~\cite{matsko2013TimingJitterMode}. Equally, the NDR representation clearly shows  the
 comb-line dependent frequency noise multiplication mechanism induced by the repetition rate variation~\cite{anderson2021photonic,lei2022OpticalLinewidthSoliton} (i.e. phase noise multiplication).  In the presence of an AMX, the  laser detuning dependence of the DKS's group velocity can exhibit the opposite sign as shown in Fig.~\ref{fig:pics1}f. Represented in the NDR, the soliton forms a similar butterfly shape with enhanced photon occupancy at the displaced mode (Fig.~\ref{fig:pics1}g).
These two effects, combined together, can then counteract due to the opposite dependence, resulting in the reduction of the detuning noise transfer (Fig.~\ref{fig:pics1}h,i).
In the following, we identify a QP by reconstructing the group velocity manifold $v_g$ as a function of $\delta\omega$ and parameters of the AMX i.e.,  mode index $\bar{\mu}$ and its displacement from the unperturbed dispersion profile $\Delta_{\bar{\mu}}$, further computing its extrema along the laser detuning direction. 
In general, this problem does not have an analytical solution, but we can efficiently reconstruct the desired dependence using a semi-analytic approach based on the Newton-Raphson method, which we describe in detail below.

\subsection{Mean-field model for noise transduction}

We model the DKS dynamics in the microresonator using the LLE with a modified dispersion profile and a Raman scattering term. In the normalized units, the generalized LLE takes the form
\begin{align}
    \frac{\partial \psi}{\partial t} 
     &= -(1+i\zeta_0)\psi + \frac{i}{2} \partial_\theta^2\psi +i|\psi|^2\psi + f \nonumber \\
     &+ v_g \partial_{\theta}\psi-i\Delta_{\bar{\mu}}        \psi_{\bar{\mu}}e^{i{\bar{m}}\theta} -i\tau\psi\partial_\theta|\psi|^2.
     \label{eq:main_LLE}
\end{align}
Here we use a standard normalization of the LLE (see Methods and Ref.~\cite{herr2014temporal} for further details) which rescales fast time (intracavity angle $\phi$) term as $\theta=\sqrt{\kappa/2D_2}\varphi$. In this way, the mode index becomes non-integer $m = \mu\sqrt{2D_2/\kappa}$, where $\mu$ --- is an integer mode index,  $\kappa$ is the total loss rate, and $D_2$ is the second-order integrated dispersion.
Terms in the second line of Eq.~\ref{eq:main_LLE}, that extend the conventional form of the LLE, represent group velocity change $v_g$, modification of the $D_\mathrm{int}$ by the AMX at the mode $\bar{\mu}$, and the Raman scattering, respectively. This formulation of the LLE is essential for the stationary solution search with the Newton-Raphson algorithm, as explained in Methods.

\subsection{QP achieved with a single-mode displacement}
To investigate the DKS group velocity response to the detuning variations $\zeta_0$, we look for a single-soliton equilibrium solution $\psi_\mathrm{DKS}$ and its relative group velocity $v_g$ for  the parameter set $(f,\zeta_0,\bar{\mu},\Delta_{\bar{\mu}})$ using the Newton-Raphson approach for Eq.~(\ref{eq:main_LLE}) described in details in Appendix~\ref{appendixA}.
To reduce the dimensionality of the parameter space, we fix the pump power $f^2=6$, $|\bar{\mu}|=21$, and sweep the detuning value within the soliton existence range (given by $\pi^2f^2/8$~\cite{herr2014temporal} in the unperturbed case) and the $\Delta_{\bar{\mu}}$ in the vicinity of the dispersive wave resonance. As a result, we obtain a soliton solution and the corresponding group velocity $v_g$  for every point on the $(\Delta_{\bar{\mu}},\zeta_0)$-subspace, both for blue- ($\bar{\mu} < 0$) and red-side ($\bar{\mu} > 0$) mode shifts (cf. Fig.~\ref{fig:pics2}).

First, we focus on the blue-sided displacement (i.e. $\bar{\mu}<0$). The presence of the shifted mode results in the generation of the DW (see Fig.~\ref{fig:pics2}a), whose strength depends on the phase matching condition with the DKS. The acquired group velocity due to the recoil increases in the vicinity of the DW resonance as shown in Fig.~\ref{fig:pics2}b. After a given value of detuning, the DW destabilizes the DKS and the equilibrium state cannot be achieved anymore (the absence of the soliton solution is depicted in black in Fig.~\ref{fig:pics2}b). 

Increasing the normalized mode shift strength $\Delta_{\bar{\mu}}$, we observe that the effect of the DW on the soliton is substantially reduced and the soliton existence range approaches the value estimated for the unperturbed LLE (see white dashed line in Fig.~\ref{fig:pics2}b). Next, we compute the group velocity directional derivative $\partial_{\zeta_0} v_g$ shown in Fig.~\ref{fig:pics2}c. As a result, we observed a family of solutions with  $\partial_{\zeta_0} v_g=0$ that correspond to the QPs. Crucially, due to the lack of control, prior experimental works have reported only a slice (vertical line cut) of the map for the fixed $\Delta_{\bar{\mu}}$ as depicted in Fig.~\ref{fig:pics2}d. Dashed lines represent the interpolated group velocity as a function of detuning $\zeta_0$ for two values $\Delta_{\bar{\mu}}=-6.85, -5$ while the solid lines represent the response $\partial_{\zeta_0} v_g$ on a logarithmic scale (directly reflecting the noise transduction). For both values $\Delta_{\bar{\mu}}$ there are two points with zero derivative $\partial_{\zeta_0} v_g$.
We followed the same procedure for the red-side mode displacement $\bar{\mu}>0$ (same side as for the Raman frequency shift) and observed similar behavior for the soliton states, group velocity, and its derivative  (Fig.~\ref{fig:pics2}(e-h)). Qualitatively, the soliton profile and its existence range remain the same as in the previous case, but the QP line is shifted now towards the higher mode-displacement amplitudes where the soliton existence range is narrowed. 

\subsection{Two-mode displacement for enhanced QP engineering}
\begin{figure*}
    \centering\includegraphics[width=\linewidth]{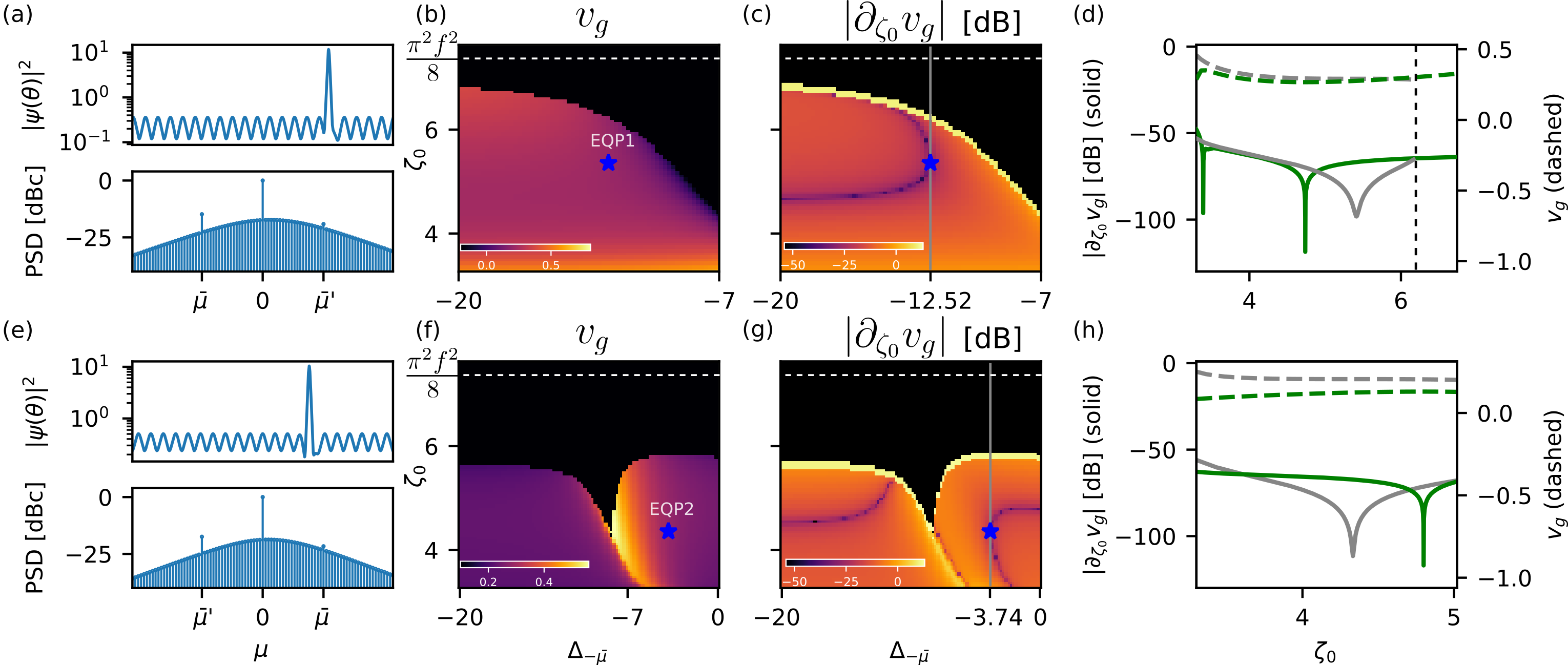}
    \caption{\textbf{Quiet point engineering} (a-d) Newton-Raphson simulations for a dispersion profile with two modes displaced: $\bar{\mu}$ and $\bar{\mu}'= -\bar{\mu}$, i.e. $(\Delta_{-\bar{\mu}},\zeta_0;\bar{\mu} = -21,\Delta_{\bar{\mu}} = -5, f^2 = 6, \tau = 5\cdot 10^{-3})$.  (a) Equilibrium DKS solution for the EQP, i.e. $(\Delta_{-\bar{\mu}}= -12.52, \zeta_0 = 5.41)$ (blue star).
    (b) Value of the group velocity in the subspace of parameters $(\Delta_{-\bar{\mu}},\zeta_0; f^2, \tau,\Delta_{\bar{\mu}})$ and (c) its  detuning directional derivative $(10 \mathrm{log}_{10}|\frac{\partial v_g}{\partial \zeta_0}|)$.
    (d) Comparison of detuning response between single mode displaced quiet point QP1 (green line, see Fig. 2.c) and engineered quiet point (EQP1) with second mode displaced (gray line in subplot c). 
    (e-h) Similar plots for the case for $\bar{\mu} = 21$, $\Delta_{\bar{\mu}} = -10.7$, $\bar{\mu}' = -21$, $\Delta_{-\bar{\mu}}= -3.74$} 
    \label{fig:pics3}
\end{figure*}

Next, we investigate the region where the two QPs (for displaced modes on the blue and the red side of the pump) can co-exist and interact. 
First, as an example of the novel dynamics, we fix the displaced mode index $\bar{\mu}=-21$ and the displacement strength $\Delta_{\bar{\mu}}=-5.00$ scanning the displacement $\Delta_{-\bar{\mu}}$, of the mode $\bar{\mu}' = -\bar{\mu}$ for different detunings $\zeta_0$. The Newton-Raphson results for the single soliton state are shown in Fig.~\ref{fig:pics3}(a-d). As in the case of a single-mode displacement, the DKS coexists with a single-period DW background (the periodicity is given by $|\bar{\mu}|$). In this case, we discovered that the single soliton solution exists for $\Delta_{-\bar{\mu}}<-7$. For large negative displacements ($\Delta_{-\bar{\mu}}\approx -20$), $\bar{\mu}'$ is out of resonance and the QP detuning value corresponds to the one in Fig.~\ref{fig:pics2}c. Reducing the displacement $|\Delta_{-\bar{\mu}}|$, the soliton starts being resonant also to mode $\bar{\mu}'$ resulting in an effective bending of the QP line, converging to the single mode one for red-shifted mode (Fig.~\ref{fig:pics2}.f). While in the case of a single-mode displacement, the QP line is always tilted (see Fig~\ref{fig:pics2}c,g) which narrows down the noise suppression region for a fixed value of $\Delta_{\bar{\mu}}$, displacing two modes, we are crucially able to \emph{engineer a flat susceptibility} over a wide range of laser detunings $\zeta_0$. We refer to this state of the system as engineered QP (EQP). The flat susceptibility region is achieved at $\Delta_{-\bar{\mu}}=-12.52$ (cf. Fig.~\ref{fig:pics3}c). In Fig.~\ref{fig:pics3}d, we compare $v_g$ and its susceptibility $\partial_{\zeta_0} v_g$ for the single mode displacement (green lines in Fig.~\ref{fig:pics2}d) with the case in Fig.~\ref{fig:pics3}c (gray lines). The latter clearly shows a flatter response profile that can be practically beneficial for accessing the QP regime. The effect of this is depicted in Figure 3 d, which shows an order of magnitude broadening of the QP detuning bandwidth. 
 We repeat the same procedure, fixing the mode $\bar{\mu}=21$ with $\Delta_{\bar{\mu}}=-10.7$ and shifting the mode $\bar{\mu}' = - \bar{\mu}$ by $\Delta_{-\bar{\mu}}$. Simulation results in Fig.~\ref{fig:pics3}(e-h) show qualitatively similar behavior, with shorter DKS existence range (defined by the fixed mode $\bar{\mu}=21$, see Fig.~\ref{fig:pics2}f). However, in this case, we find two QP families for a single DKS solution. The flattest response is achieved at $\Delta_{-\bar{\mu}}=-3.74$ (see Fig.~\ref{fig:pics3}(g,h)). In this way, we observe that careful control over the two-mode displacement can extend the noise suppression region of the QP in the parameter space. 

\subsection{Linear stability of the solutions}
\begin{figure*}
    \centering
    \includegraphics[width=\linewidth]{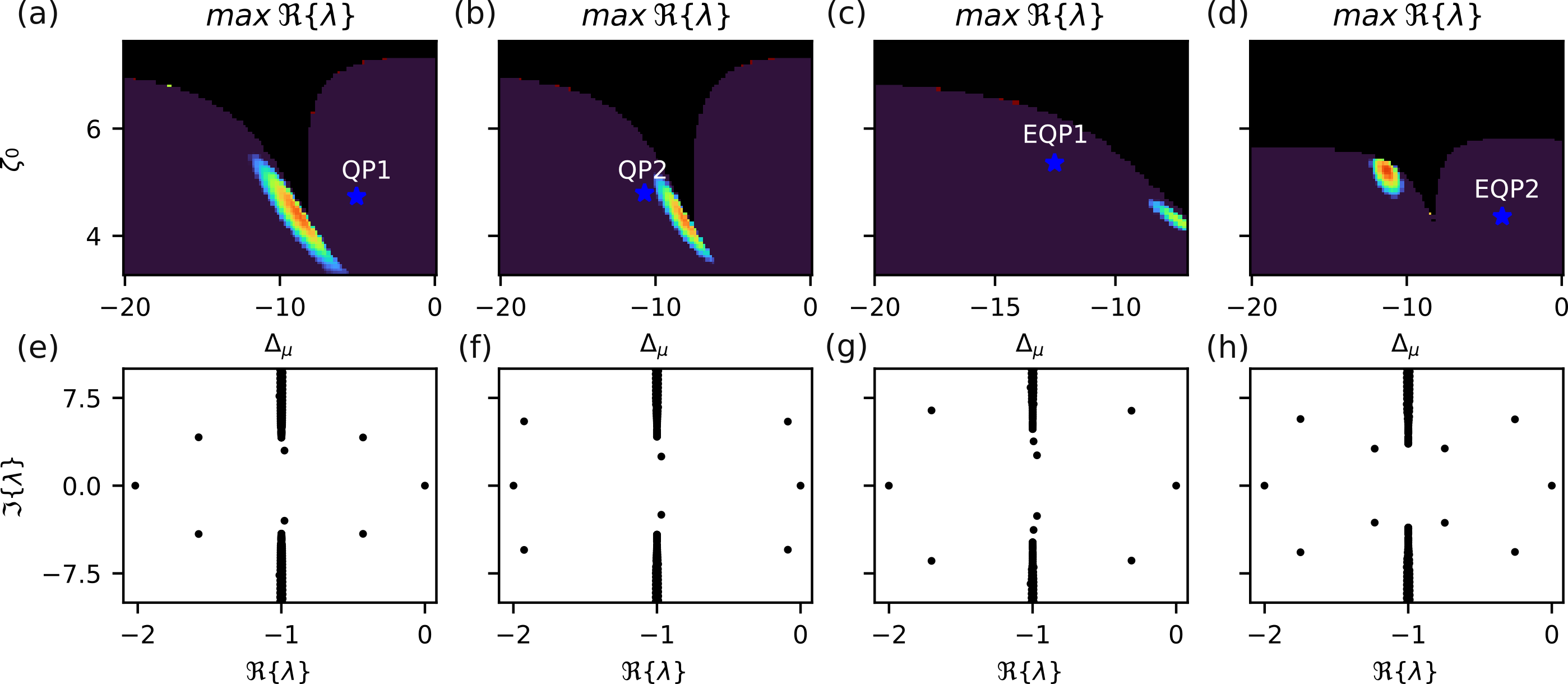}
    \caption{\textbf{Linear stability of the quiet points} (a-d) Value of the real part of the Jacobian eigenvalue ($\lambda$) with highest real part, in the parameters subspaces considered in fig. 2.b,2.f,3.b,3.f respectively. The analysis shows there are regions within the DKS existence range in which the soliton solutions are linearly unstable (colored regions) but those do not include the analyzed QP that is consequently considered stable.
    (e-h) Detail on the spectrum of the Jacobian for the cases in (a-d). The spectrum shows that the instabilities arising in the colored regions are due to the Hopf bifurcation, considered the presence of pairs of conjugated soft modes approaching $\Re\{\lambda\}=0$.}
    \label{fig:pics4}
\end{figure*}
Having derived above the equilibrium value of the soliton group velocity and its relative soliton state, for different mode shifts and having observed QP lines in various configurations using the Newton-Raphson method, a crucial question arises that we answer next:  \emph{are the observed states stable and/or is there any additional instability region?} To estimate the stability of the equilibrium solutions, we perform linear stability analysis, numerically investigating the eigenvalues $\lambda$ of the Jacobian operator associated with~\eqref{eq:main_LLE} for each particular soliton state found in the previous section (see Methods). In particular, we focus on the eigenvalues with the greatest real part, since, if positive, those are responsible for the linear growth of any perturbation around the equilibrium. The real part of the latter ($\max \Re\{\lambda\}$) are plotted in Fig.\ref{fig:pics4}.a-d. We observe that the soliton solutions are linearly stable almost everywhere in the considered subspace, and in particular at the QPs. Exceptions are for a narrow region in correspondence with the reduced existence range (reminding an instability tongue). In those regions there exist at least one eigenvalue with positive real part. From the actual structure of the spectrum of the Jacobian, computed for the quiet point states, (Fig.\ref{fig:pics4}.e-h) we find that these instabilities are due to Hopf bifurcations, characterized by the vanishing real parts of a pair of the complex conjugated soft modes. In this region of parameter space, those will be responsible for the transition from stable solitons to breathing states.
\subsection{Dynamical simulation of the phase noise transfer}

\begin{figure*}
    \centering
    \includegraphics[width=\linewidth]{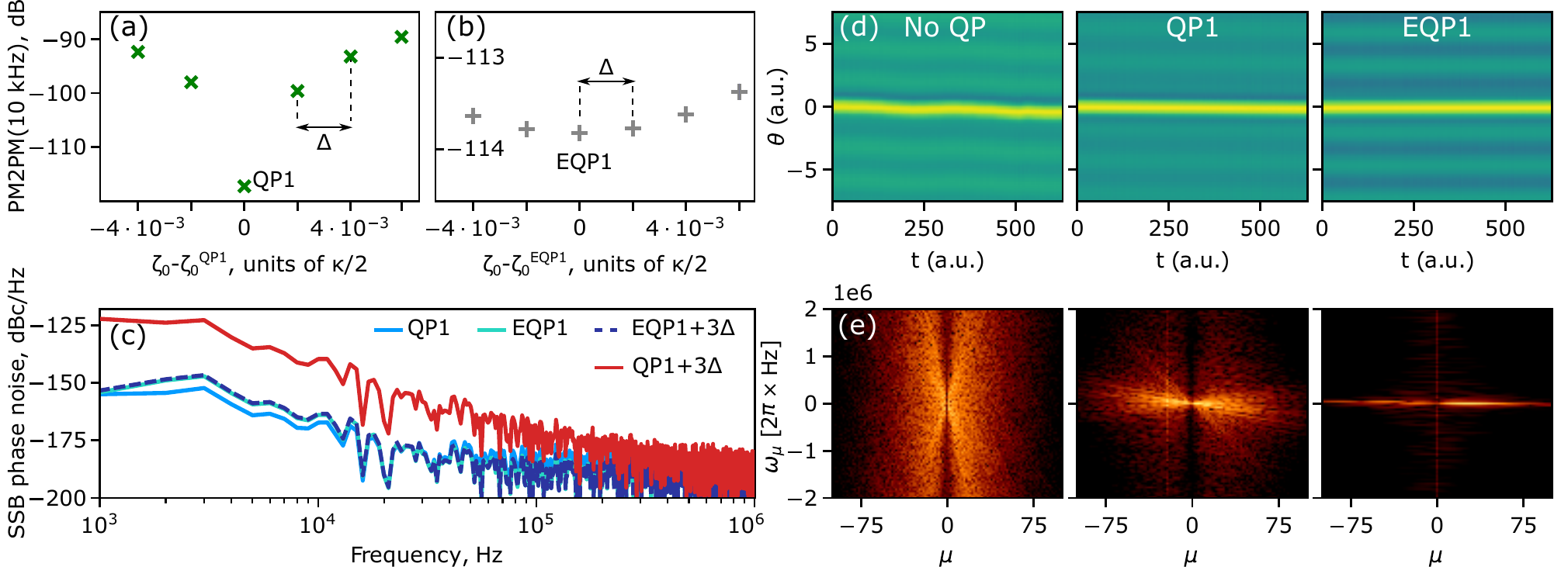}
    \caption{\textbf{Dynamical simulations of the noise transfer for different operating points.}  
     PM2PM coefficient (ratio between the microwave phase noise and the pump laser phase noise) at 10 kHz modulation frequency in the vicinity of QP1 (green crosses in panel a) and EQP1 (gray crosses in panel b). (c) SSB phase noise plot for exact laser detuning corresponding to QP1 and EQP1 as well as a deviation of $3\Delta = 3 \cdot 10^{-3}$ $\kappa$.
    (d) Spatiotemporal diagram of the DKS evolution under the influence of amplified pump phase noise. (e) Nonlinear dispersion relation, for different operating points.  The noise transduction to the soliton repetition rate is greatly reduced in the presence of two engineered mode-crossings. }   
    \label{fig:pics5}
\end{figure*}

To compare the phase noise performance of different operating points, the dynamical evolution has been simulated with the  step-adaptative Dormand-Prince Runge-Kutta method of Order 8(5,3)~\cite{press2007numerical}. We perform the direct dynamical simulations of the LLE adding a realistic noise to the detuning term measured experimentally from the Topical CTL 1550 laser having a standard deviation of 5~kHz (see Methods). In this way, we simulate two DKS operating points: QP1 (see Fig.~\ref{fig:pics2}.b) and EQP1 (see Fig.~\ref{fig:pics3}b). 
The difference between the phase noise transfer performance at different operating points can be clearly seen in Fig.~\ref{fig:pics5}a,b.  A series of numerical experiments confirm the conclusion obtained with the stationary solver analysis. Indeed, changing the central frequency of the pump by the value of 10$^{-3}$~$\kappa$, we clearly observe that the overall noise transduction from the pump to the soliton repetition rate (PM2PM at 10 kHz offset) performance of EQP1 increases by 0.5~dB for $3 \cdot 10^{-3}$~$\kappa$, while in the case of QP$_1$, we observe $> 28$~dB of the transduction enhancement. Corresponding single-sideband (SSB)  phase noise performance is depicted in Fig.~\ref{fig:pics5}c. As one can see, the fluctuations of the central frequency of the pump laser do not visibly affect the performance of the system at EQP1.

Next, we verify that in the case of the noisy pump lasers (standard deviation of 8\% $\kappa/2$) EPQ leads to a significant noise reduction due to the larger noise suppression region. We employed the same phase noise data, scaling it to obtain the detuning fluctuation of the order of the width of the standard QP, i.e. 8\% $\kappa/2$. In the parameter regime outside of the QP region the influence of the pump fluctuations on the DKS dynamics is visible directly from the spatiotemporal diagram (see Fig.~\ref{fig:pics5}d). To distinguish the performance of QP1 and EQP1, we can use the NDR to represent its effect (Fig.~\ref{fig:pics5}e). We observe a clear suppression of the noise multiplication in the case of EQP1. This confirms our predictions based on the group velocity variation obtained with the stationary solution solver. 

\section{Conclusions and discussion}
In summary, we have demonstrated a method to increase the effectiveness of QPs, that are central to achieving low phase noise soliton microcombs for microwave generation. Our work shows that engineering QP introduced via two dedicated and controllable mode crossings enables one to create broader regions of enhanced noise suppression. Our work is directly implementable using current technology and provides a new approach to the enduring challenge of obtaining thermal noise-limited micro-wave generation from integrated soliton microcombs, which in contrast to crystalline resonators employ materials such as silicon nitride or silica, that do exhibit a Raman self-frequency shift.

These results were obtained via a semi-analytical approach, based on the Newton-Raphson method, studied the phenomenology of QP in the presence of Raman scattering, dispersive waves, and detuning noise, within a simplified model of AMX.  This allowed us to obtain several insights  (i) QPs can be achieved by placing AMX on both blue- and red-detuned sides of the pump. This highlights the fact that not the absolute value of the frequency shift must be compensated, but its derivative over the laser detuning. (ii) Engineering the interaction of two QPs leads to further reduction of the noise transfer. (iii) The EQPs predicted in this work are linearly stable and characterized by more than 28 dB reduction of the PM2PM coefficient with respect to a generic QP described in the literature when a detuning deviation of the order of $0.03\%$ of $\kappa$ is introduced. 
We anticipate that the detuning-dependent variation of the repetition rate can be completely eliminated by further controlling the integrated dispersion profile for example by corrugating the microresonator circumference~\cite{lucas2022tailoring}, which is however outside the scope of this work. 


\begin{acknowledgments}
 This material is based upon work supported by the Air Force Office of Scientific Research under award number FA9550-19-1-0250. The authors thank Savyaraj Deshmukh for the critical reading of the manuscript.
\end{acknowledgments}


\section{Appendix A: Mean field model}

The modified LLE, presented in the main text, which accounts for the Raman scattering and effect induced by the AMX:
\begin{align}
    \frac{\partial \psi}{\partial t} 
     &= -(1+i\zeta_0)\psi + \frac{i}{2} \partial_\theta^2\psi +i|\psi|^2\psi + f \nonumber \\
     &+ v_g \partial_{\theta}\psi-i\Delta_{\bar{\mu}}        \psi_{\bar{\mu}}e^{i{\bar{m}}\theta} -i\tau\psi\partial_\theta|\psi|^2 ,
\end{align}
where $t=t'/2\tau_{\mathrm{ph}}$ is the time normalized to photon lifetime $\tau_\mathrm{ph}=1/\kappa$, $\kappa=\kappa_0+\kappa_{\mathrm{ex}}$ is the total linewidth of the cavity composed of the internal linewidth $\kappa_0$ and the coupling to the bus waveguide $\kappa_\mathrm{ex}$. The normalized laser-cavity detuning is $\zeta_0=2\delta\omega/\kappa$, the fast time is defined as $\theta=\sqrt{\kappa/2D_2}\varphi$ with $D_2$ as the GVD and $\varphi \in [-\pi,\pi]$ as the azimuthal coordinate inside the cavity. $\bar{\mu}$ indicates the shifted mode number and $\psi_{\bar{\mu}}$ is the amplitude of the displaced mode. We point out that, due to the normalization of the fast time coordinate, the mode numbers $m$ are not integer, however, they are related to the actual comb line index $\mu$ by a simple multiplication factor, i.e. $m = \mu\sqrt{2D_2/\kappa}$.  Thus, referring to an integer $\Bar{\mu}$, we imply the comb line index $\Bar{\mu}$ is associated with. The normalized pump power is $f=\sqrt{8\kappa_\mathrm{ex}g_0/\kappa^3}s_\mathrm{in}$ where $g_{0}$ is single photon Kerr frequency shift and $s_\mathrm{in}=\sqrt{P_\mathrm{in}/\hbar\omega}$ and where 
$|s_\mathrm{in}|^2$ is the laser photon flux. The last two terms in Eq.~(\ref{eq:main_LLE}) describe the single-mode AMX and the Raman scattering respectively. The normalized mode displacement is defined as  $ \Delta_{\bar{\mu}}\delta_{\mu, \bar{\mu}} = 2(D_\mathrm{int}(\mu)-D_2 \mu^2/2)/\kappa $, with $\delta_{\mu,\bar{\mu}}$ as Kronecker delta, and the normalized Raman shock time is $\tau=\tau_\mathrm{R}D_1\sqrt{\kappa/2D_2}$.

The single-mode AMX term comes directly from the modified dispersion profile (e.g., Fig.~\ref{fig:pics1}b) as follows from 
\begin{equation*}
      \mathcal{F}^{-1}\Big[(\frac{\mu^2}{2} + \Delta_{\bar{\mu}}\delta_{\mu,\bar{\mu}}) \psi_\mu \Big] = -\frac{1}{2} \partial^2_{\theta}\psi +\Delta_{\bar{\mu}}        \psi_{\bar{\mu}}e^{i{\bar{m}}\theta},
\end{equation*}
where $\mathcal{F}^{-1}[...]$ stands for the inverse Fourier transform.
We simulate a system with the following parameters: $\kappa/(2D_2) = 76.923$, $\tau_\mathrm{R} D_1 = 0.0006$.

\section{Appendix B: Newton-Raphson method for the QP}
\label{appendixA}
The stationary solitons state and their respective value of group velocity have been computed from Eq.~\eqref{eq:main_LLE} applying the Newton-Raphson method. This method is used to compute the solutions of a nonlinear (in principle vectorial) equation of the form of 
\begin{equation}
    \mathbf{F}(\mathbf{\varPhi}) = 0,
    \label{NR-simple}
\end{equation}
by applying the following iterative scheme: 
\begin{equation}
    \begin{cases}
        \mathbf{\varPhi}_{k+1} = \mathbf{\varPhi}_{k} - \underline{\underline{J}}^{-1}(\mathbf{\varPhi}_{k})\mathbf{F}(\mathbf{\varPhi}_{k}) \\
        \mathbf{\varPhi}_0: \mathrm{Initial\, guess}
        \label{NR-algorithm}
    \end{cases}
\end{equation}
Where $\underline{\underline{J}}^{-1}$ is the inverse of the Jacobian matrix of the function $\mathbf{F}$~\cite{kelley2003SolvingNonlinearEquations,seydel2010PracticalBifurcationStability}, i.e.
\begin{equation}
    J_{i,j} = \frac{\partial F_i}{\partial \mathbf{\varPhi}_j} 
\end{equation}
If the initial condition has been chosen correctly and the Jacobian remains invertible, the algorithm will converge to the desired solution $\mathbf{\varPhi}^{\star}$, fixed point of \eqref{NR-algorithm}, i.e. 
\begin{equation}
    \mathbf{\varPhi}^{\star} = \mathbf{\varPhi}^{\star}- \underline{\underline{J}}^{-1}(\mathbf{\varPhi}^{\star})\mathbf{F}(\mathbf{\varPhi}^{\star}) 
    \label{fixed-point}
\end{equation} and so characterised by $\mathbf{F}(\mathbf{\varPhi}^{\star}) = 0$.

In our case, we exploit the method to find stationary single soliton solutions, $\psi$, of Eq.~\eqref{eq:main_LLE} and the corresponding group velocity $v_g$ by adding it explicitly as an optimization variable.
In this case equation \eqref{NR-simple} is set as the following:
\begin{equation}
     \mathbf{F}(\psi,\psi^*,v_g) := \begin{bmatrix}
         g(\psi,\psi^*,v_g)\\
    g^*(\psi,\psi^*,v_g)\\
    \Re\{\partial_\theta\psi\}|_{\theta = \theta_{max}}
     \end{bmatrix}=0
     \label{actual_NR}
 \end{equation} 
  where $g(\psi,\psi^*,v_g)$ is the r.h.s of equation \eqref{eq:main_LLE}. 
 To implement the iterative algorithm from Eq.~\eqref{NR-algorithm} for equation~\eqref{actual_NR}, we define $\mathbf{\Psi} := (\psi,\psi^*,v_g)$ and rewrite the function $\mathbf{F}(\mathbf{\Psi})$ as a formal matrix product
 \begin{equation}
     \mathbf{F}(\mathbf\Psi) = \underline{\underline{\Tilde{\operatorname{F}}}} (\mathbf{\Psi})\mathbf{\Psi} \equiv\begin{bmatrix}
    \Tilde{\operatorname{F}}_{1,1} & \Tilde{\operatorname{F}}_{1,2}& \Tilde{\operatorname{F}}_{1,3}\\
    \Tilde{\operatorname{F}}_{2,1}& \Tilde{\operatorname{F}}_{2,2} & \Tilde{\operatorname{F}}_{2,3}\\
    \Tilde{\operatorname{F}}_{3,1}& \Tilde{\operatorname{F}}_{3,2}& \Tilde{\operatorname{F}}_{3,3}\\
    \end{bmatrix} \begin{bmatrix}
        \psi \\ \psi^*\\ v_g
    \end{bmatrix}  
 \end{equation}
where $\underline{\underline{\Tilde{\operatorname{F}}}} (\mathbf{\varPsi} )$ is a $3\times3$ matricial operator defined by the following equations 
:\\
\begin{gather*}
    \Tilde{\operatorname{F}}_{1,1} = -1-i(\zeta_0 - \frac{1}{2}\partial_\theta^2 + \Delta_{\bar\mu} e^{i\bar\mu \theta }\hat{\mathcal{F}}_{\bar{\mu}})\\
    \Tilde{\operatorname{F}}_{1,2} = i\psi^2(1 - \tau \partial_\theta) - i\tau\psi\partial_\theta\psi \\
    \Tilde{\operatorname{F}}_{1,3} = \Tilde{\operatorname{F}}_{2,3} = \partial_\theta\\
    \Tilde{\operatorname{F}}_{2,2} = (\Tilde{\operatorname{F}}_{1,1})^*\\
    \Tilde{\operatorname{F}}_{2,1} = (\Tilde{\operatorname{F}}_{1,2})^*\\
    \Tilde{\operatorname{F}}_{3,1} = \Tilde{\operatorname{F}}_{3,2} =  \frac{1}{2}\int_{\mathcal{R}}d\theta \delta(\theta - \theta_{max})\partial_{\theta}\\
     \Tilde{\operatorname{F}}_{3,3} = 0\\
     \hat{\mathcal{F}}\psi = \int d\theta \psi e^{-i\mu\theta} = \psi_{\mu}\\
     \hat{\mathcal{F}}^{-1}\psi_{\mu} = \sum_{\mu} \psi_{\mu} e^{i \mu \theta} = \psi\\
     \hat{\mathcal{F}}_{\bar{\mu}}\psi = \sum_{\mu}\delta_{\mu,\bar{\mu}}\hat{\mathcal{F}}\psi = \psi_{\bar{\mu}}
    \label{eq:matrix_operator}
\end{gather*}
In this way, the iterative equation in \eqref{NR-algorithm} can be rewritten as 
\begin{equation}
     \underline{\underline{J}}\mathbf{\Psi}_{k+1} = (\underline{\underline{J}} - \underline{\underline{\Tilde{\operatorname{F}}}}) \mathbf{\Psi}_{k} 
\end{equation}
that can be numerically solved for $\mathbf{\Psi}_{k+1}$. 

Thus, the Jacobian in the rotating frame takes form
\begin{gather}
    \underline{\underline{{\operatorname{J}}}}( \mathbf{\Psi})= \underline{\underline{\Tilde{\operatorname{F}}}}( \mathbf{\Psi}) +  \begin{bmatrix}
    \setlength{\tabcolsep}{3.5pt}
    \hat{\Delta}( \mathbf{\Psi}) & 0 & 0 \\
    0 & \hat{\Delta}^*( \mathbf{\Psi}) &0\\
    0&0&0
    \end{bmatrix}
    \label{eq:jacobian_matrix}
\end{gather}
where:
\begin{equation}
\hat{\Delta}(\Psi) := 2i|\psi|^2+v_g\partial_\theta -i\tau(\partial_{\theta} |{\psi}|^2 + |\psi|^2\partial_\theta  + \psi\partial_\theta \psi^*).
\end{equation}
We implement the differential operators appearing in the matrices entries using the  discrete Fourier transform matrix (DFT matrix, \cite{Winograd1978-wj}), which allows us to express those operators in the Fourier space where they are represented algebraically.

To control the numeracal convergence of the algorithm, we use the standard measure
\begin{equation}
    \sqrt{\frac{\lVert\mathbf{\Psi}_{k+1} - \mathbf{\Psi}_{k}\rVert^2_2}{\lVert\mathbf{\Psi}_{k}\rVert^2_2}} < 10^{-6} 
\end{equation}
where $\lVert\cdot \rVert_2$ is the ${L}^2$-norm.
To avoid discretization problems, (especially for the computation of the correct spectrum of the Jacobian), we discretized the envelop function $\psi$ in $N_{\psi}=2^{10}$ number of points.

Finally, some comment on the choice of the initial condition $\Psi_{0}$ is required.
Despite its power, this method, being the simplest of its kind, suffers from a small convergence radius. That means the initial condition must be already very close to the actual one. To overcome the problem,
we used a numerical DKS solution from dynamical simulations as a guess solution with zero group velocity. For the subsequent points, we used instead as seeds the converged solution from the closest point in the detuning direction.



\section{Appendix C: Dynamical simulation of the noise transduction}

The dynamical simulations have been carried out with a  the step-adaptative Dormand-Prince Runge-Kutta method of Order 8(5,3)~\cite{press2007numerical}, hard seeded an approximate DKS solution. The input pump phase noise has been obtained from a linearization of the data of the power spectral density data of Toptica CTL 1550 Laser, $S^{\mathrm{in}}_{\phi}$. In particular, it has been implemented through a detuning noise term $\zeta_0(t)$ obtained as  
\begin{equation}
    \zeta_0(t) = \alpha \hat{\mathcal{F}_{t}}( \sqrt{\nu^2 S^{\mathrm{in}}_{\phi}(\nu)} e^{i x(\nu)} ) 
\end{equation}
where a uniformly distributed random phase $x(\nu)$ has been added to each frequency to obtain a random realization of the detuning noise and coefficient $\alpha $ normalizes the standard deviation of the pump detuning.  


Output field  in real units for the case of critical coupling ($\kappa_\mathrm{ex} = \kappa_\mathrm{0}$):
\begin{equation}
    P_{\mathrm{out}}(\phi,t) = \hbar \omega_0\frac{\kappa_{ex}^2}{g_0}|f - \psi|^2
\end{equation}
Spectrogram:
\begin{equation}
    P_{\mu}(t) = \hat{\mathcal{F}}_{\mu}P_{\mathrm{out}}
\end{equation}
where, $\hat{\mathcal{F}}_{\mu}$ is the operator taking the $\mu$-th Fourier component of a function (the power in this case), as defined in the previous section.
The phase of the first comb line (repetition rate phase):
\begin{equation}
    \phi(t) = \mathrm{arg}\big[P_{1}(t)\big],
\end{equation}
where $\mathrm{arg}\big[P_{1}(t)\big]$ denotes the phase of the first complex Fourier component of the detected optical power.
The spectrum of phase noise:
\begin{equation}
    S_{\phi}(\nu) = |\hat{\mathcal{F}}_{\nu}\phi(t)|^2
\end{equation}
The transduction coefficient has been computed as:
\begin{equation}
    \mathrm{PM2PM} = 10log_{10} \frac{S_{\phi}}{S^{\mathrm{in}}_{\phi}}
\end{equation}
We point out in this analysis we assume an ideal photodetector, neglecting so its actual response function. 
As for the Newton-Raphson method (see Appendix~\ref{appendixA}), we discretized the fast time axis (i.e. azimuthal coordinate) in $N_{\psi}= 2^{10}$ points while the slow time  in $N_{t} = 20000$ points. In addition, in order to obtain a sensitivity of the order of the kHz, we simulate the soliton dynamics for 1 ms.

\end{document}